\documentclass{article}
\usepackage{times,color,amsfonts,amssymb,amsbsy,amsthm,amsmath,graphics,graphicx,hyperref,bm,bbm,mathrsfs,color,xcolor,cancel,float,MnSymbol,bbold}
\usepackage[normalem]{ulem}
\usepackage{authblk}
\usepackage[section]{placeins}
\hypersetup{colorlinks,linkcolor={blue},citecolor={blue},urlcolor= {blue}}

\newcommand{\ignore}[1]{}
\newtheorem{remark}{Remark}

\let\oldsqrt\sqrt
\def\sqrt{\mathpalette\DHLhksqrt}
\def\DHLhksqrt#1#2{%
\setbox0=\hbox{$#1\oldsqrt{#2\,}$}\dimen0=\ht0
\advance\dimen0-0.2\ht0
\setbox2=\hbox{\vrule height\ht0 depth -\dimen0}%
{\box0\lower0.4pt\box2}}

\DeclareFontFamily{OT1}{pzc}{}
\DeclareFontShape{OT1}{pzc}{m}{it}%
              {<-> s * [1.25] pzcmi7t}{}
\DeclareMathAlphabet{\mathpzc}{OT1}{pzc}%
                                 {m}{it}

\newcommand{\Rea}{\mathbb{R}}

\newcommand{\Ex}{\mathbb{E}}

\newcommand{\Int}{\mathbb{Z}}

\begin{document}

\title{A Thermodynamical derivation of the quantum potential and the temperature of the wave function}
\author{Curcuraci L.$^{1,2}$,  Ramezani M $^{3,4}$}
\date{%
	$^1$\small{\emph{Department of Physics, University of Trieste, Strada Costiera 11 34151, Trieste, Italy}}\\%
	$^2$\small{\emph{Istituto Nazionale di Fisica Nucleare, Trieste Section, Via Valerio 2 34127, Trieste, Italy}}\\[2ex]%
	$^3$ \small{\emph{Department of Physics, Sharif University of Technology, Tehran 14588, Iran}}\\
	$^4$ \small{\emph{School of Physics, Institute for Research in Fundamental Sciences (IPM), Tehran 19395, Iran}}
	\today
}
\maketitle

\begin{abstract}
In this paper a thermodynamical derivation of the quantum potential is proposed. Within the framework of Bohmian mechanics we show how the quantum potential can be derived, by adding an additional informational degree of freedom to the ordinary degrees of freedom of a physical system. Such a derivation uses the First Law of thermodynamics for this additional degree of freedom and basic equilibrium thermodynamics methods. By doing that, one may associate a temperature to each wave function. Features and behavior of this temperature in different situations is studied.
\end{abstract}

\section{Introduction.}\label{sec1}

In quantum mechanics, a particle moving in a 3-dimensional space is described by a wave function $\psi(\mathbf{x},t)$. By writing $\psi(\mathbf{x},t)$ in polar form 
\begin{equation}\label{polar}
\psi(\mathbf{x},t)=\sqrt{p(\mathbf{x},t)}\exp\left(\frac{i}{\hbar}S(\mathbf{x},t)\right),
\end{equation}
one can define a key quantity, useful in many causal interpretations of quantum mechanics: the \emph{quantum potential}
\begin{equation}\label{quantumpotential}
\begin{split}
		Q(\mathbf{x},t) &:= -\frac{\hbar^2}{4m} \frac{\nabla^2\sqrt{p(\mathbf{x},t)}}{\sqrt{p(\mathbf{x},t)}} = - \frac{\hbar^{2}}{8m}\bigg(\frac{\nabla p(\mathbf{x},t)}{p(\mathbf{x},t)}\bigg)^{2}+\frac{\hbar^{2}}{4m}\frac{\nabla^{2}p(\mathbf{x},t)}{p(\mathbf{x},t)}.
\end{split}
\end{equation}
This quantity probably appears the first time in 1927 in the work of E. Madelung \cite{1927-Madelung}, where an analogy between quantum mechanics and hydrodynamics was proposed. However, its role in the explanation of quantum phenomena was recognized only after the seminal work of D. Bohm. In \cite{1952-Bohm1,1952-Bohm2}, he proposed a consistent theory of point-like particles in motion which reproduces the results of quantum mechanics. A central ingredient of such a theory is the quantum potential \eqref{quantumpotential}. Indeed, it was proved that typical quantum phenomena like interference in a double-slit experiment  \cite{1979-Philippidis}, or the quantum tunneling effect \cite{1982-Dewdney}, can be consistently explained by taking into account the contribution of $Q$ to the motion of an otherwise classical point-like particle. 

The quantum nature of $Q$ can be understood by observing that the way quantum potential affect the motion of a particle does not depend on the intensity of the wave function, as opposite of what would happen for a classical wave hitting a particle. Indeed, if we double the intensity of $\psi$, $Q$ remains the same, while in the classical case the effect would be stronger. It is also interesting to observe that the quantum potential appears and plays a central role also in other formulation of quantum mechanics like in Nelson stochastic mechanics \cite{1966-Nelson}. 

Consider the Schr\"{o}dinger equation for a particle in a potential $V(\mathbf{x})$,
\begin{equation}
\label{Shro-Eq}
-i\hbar\frac{\partial}{\partial t}\psi(\mathbf{x},t) =\frac{-\hbar^2}{2m}\nabla^{2}\psi(\mathbf{x},t) + V(\mathbf{x})\,\psi(\mathbf{x},t).
\end{equation}
Inserting \eqref{polar} in the Schr\"{o}dinger equation, from its real and the imaginary part one obtains two coupled real-valued differential equations
\begin{align}
\label{continuity}
\frac{\partial}{\partial t}p(\mathbf{x},t)&=-\nabla \cdot \bigg(\frac{\nabla S(\mathbf{x},t)}{m}\,p(\mathbf{x},t)\bigg),\\
\label{Hamil-Jacob}
\frac{\partial}{\partial t}S(\mathbf{x},t)&=-\bigg(\frac{(\nabla S(\mathbf{x},t))^{2}}{2m}+V(\mathbf{x})+Q(\mathbf{x},t)\bigg).
\end{align}  
Considering $\nabla S(\mathbf{x},t)/m$ as the \emph{velocity} field in which the particle moves, equations \eqref{continuity} and \eqref{Hamil-Jacob} are respectively the well-known \emph{continuity} and \emph{Hamilton-Jacobi-Madelung} equations. It is known that these equations, supplemented with the Wallstrom condition \cite{1994-Wallstrom}, i.e.  
\begin{equation}
	\oint_\gamma \nabla S(\mathbf{x},t) \cdot d\mathbf{x} = n\hbar \mspace{50mu} n \in \Int,
\end{equation}
with $\gamma$ closed path in the configuration space of the system, are equivalent to the Schr\"{o}dinger equation. 

Already Bohm \cite{1952-Bohm1}, recognized that the two equations of motion above can be derived using variational methods from an Hamiltonian functional. Defining
\begin{equation}
\label{Hamiltonian}
\mathcal{H}[p,S] :=\int\left(\frac{(\nabla S(\mathbf{x}))^{2}}{2m}+V(\mathbf{x})+\frac{\hbar^{2}}{8m}\bigg(\frac{\nabla p(\mathbf{x})}{p(\mathbf{x})}\bigg)^{2}\right)p(\mathbf{x}) d\mathbf{x},
\end{equation}
by the ordinary principle of least action one obtains the variational equations $\dot{p}=\delta \mathcal{H}/\delta S$ and $\dot{S}=-\delta \mathcal{H}/\delta p$, reproducing exactly \eqref{continuity} and \eqref{Hamil-Jacob} respectively. We also observe that the same Hamiltonian functional is used in stochastic mechanics to derive the equations of motion \cite{1983-Guerra,1988-Guerra}. The Hamiltonian functional \eqref{Hamiltonian} has a straightforward physical interpretation: it is a sort of average energy of the particle. Specifically, from classical Hamiltonian mechanics, the first two terms can be understood as the average kinetic and potential energy of the particle. Quantum effects are related to the last term of \eqref{Hamiltonian}, i.e.
\begin{equation}
\label{QP}
U_q:= \frac{\hbar^{2}}{8m}\int\bigg(\frac{\nabla p(\mathbf{x})}{p(\mathbf{x})}\bigg)^{2}p(\mathbf{x}) d\mathbf{x}
\end{equation}
since the ordinary quantum potential is related to the previous quantity by means of the following
\begin{equation}\label{QPformula}
	Q = -\frac{\delta U_q}{\delta p}.
\end{equation}


Many attempts to derive the quantum potential $Q$ from first principles are available in literature, see \cite{2014-Licata} and references therein. In the D\"{u}rr-Goldstein-Zangh\'{i} formulation of Bohmian mechanics \cite{1992-Durr,1992-Durr2,2009-Durr}, the quantum potential does not have a central role and it emerges by taking the time derivative of the guiding equation (which is assumed to be more fundamental). A more fundamental role is assigned to $Q$ in the quasi-Newtonian approach \cite{2013-Atiq}, where the actual form \eqref{quantumpotential} of the quantum potential is derived by minimizing (a modified version of) the energy of a particle. Other derivations assumes that the quantum potential is a manifestation of a supposed non-euclidean nature of the 3-dimensional space in which the particle moves. These approaches goes under the name of geometrodynamical approaches. Among them we find the work of Fiscaletti based on Weyl geometry \cite{2014-Licata,2012-Fiscaletti}, and the attempt by Hiley to formalize Bohm's idea of \textquotedblleft implicate/explicate order" using non-commutative geometry \cite{2005-Hiley,2010-Hiley}. There are also information-based approaches. In \cite{2010-Carroll} the quantum potential is derived from the Fisher information stored in the probability distribution of the particle, while in \cite{2009-Sbitnev}  it is derived from a novel notion of quantum entropy. Finally, the quantum potential $Q$ can be derived from thermodynamical considerations, as it was done in \cite{2008-Grossing,2009-Grossing}. In this last approach, the quantum potential is considered as an additional kinetic energy a particle may absorb by a heat flow from a thermal field (quantum vacuum energy).

In all the approaches mentioned till now, starting from first principles the various method lead directly to the formula \eqref{quantumpotential} for quantum potential $Q$ directly. Here we propose a method to derive $U_q$ from first principles, from which the quantum potential can be derived using \eqref{QPformula}. This approach was also followed in \cite{1998-Recami}, but there $U_q$ is essentially due to the internal motion of a spinning particle. Here instead, it will be argued that, associating an energy for the informational degree of freedom of a particle, $U_q$ can be explained by using simple thermodynamical considerations. More precisely, we show that one can derive $U_q$ if one assumes an additional informational degree of freedom for a particle with average energy given by $\frac{1}{2}k_{B}\tilde{T}$, where $\tilde{T}$ is a temperature defined on the configuration space of the quantum system.

\section{Quantum Potential and Information}\label{sec2}

Consider the term \eqref{QP}. It was already observed in \cite{2009-Sbitnev} that $U_q$ is proportional to the Fisher information \cite{1925-Fisher} stored in the probability distribution of the particle position. Indeed, given a probability distribution $p(\mathbf{x},\theta_1,\cdots,\theta_m)$, where $\theta_1,\cdots,\theta_m$ are some unknown parameters, the Fisher information $\mathcal{I}[p]$ associated to it is defined as
\begin{equation}
\begin{split}
	\mathcal{I}[p] &:= \Ex\left[ (\nabla_{\mathbf{\theta}} \log p(\mathbf{x},\theta_1,\cdots,\theta_m))^2 \right] \\
	&= \int \sum_i(\partial_{\theta_i} \log p(\mathbf{x},\theta_1,\cdots,\theta_m))^2 p(\mathbf{x},\theta_1,\cdots,\theta_m)d\mathbf{x}.
\end{split}
\end{equation}
One can immediately see that for a particle with wave function $\psi(\mathbf{x})$, $U_q$ is proportional to $\mathcal{I}[|\psi(\mathbf{x})|^2]$, when the parameters $\theta_1,\cdots, \theta_m$  are three and coincide with the coordinates of the particle $x_1,x_2,x_3$. This is not the first time that Fisher information plays a role in quantum mechanics at fundamental level. For example, by using a variational principle by minimizing the Fisher information one can derive the Schr\"{o}digenr equation  \cite{1989-Frieden,1998-Reginatto}. On the other hand it can be shown that the Heisenberg  uncertainty principle is just the Cramer-Rao bound on the Fisher information of the position of a particle \cite{2000-Luo}. 

The reason why information should play such a fundamental role in the dynamics of a quantum particle is not so clear. In this section we develop a possible argument to motivate this fact. Following an analogy with classical mechanics, here we argue that  \eqref{QP} represents the energy associated to the informational degree of freedom of the particle. In particular, by assuming an analogous work-energy theorem for the informational degree of freedom, we can motivate the form of $U_q$.

Suppose that in some process, the position probability density function of a particle changes from $p$ to $p'$. A natural question can be: how much energy is needed for this change? To answer this question we may look at the analogue situation in classical mechanics. Consider the motion of a classical particle in a 3-dimensional space. Let $\mathbf{x}(t) = (x_1(t),x_2(t),x_3(t))$ denote the position of a particle at time $t$. The $i$-th component of the particle velocity $\mathbf{v}(t) = (v_1(t),v_2(t),v_3(t))$ at each instant of time is defined as
\begin{equation}
	v_i(t) := \lim_{h\to 0}\frac{d[x_i(t+h),x_i(t)]}{h} ,
\end{equation}
where $d[a,b]$ is the distance between the points $a$ and $b$. For a point-like particle in classical mechanics, the distance $d$ is just the euclidean distance in $\Rea^3$. According to the work-energy theorem \cite{2005-Taylor}, the energy needed to change the position of the particle from $\mathbf{x}_{1}=\mathbf{x}$ to $\mathbf{x}_{2}=\mathbf{x}+d\mathbf{x}$ is proportional to $\mathbf{v}\cdot d\mathbf{v}$, where $d\mathbf{v}$ is the change in the velocity of the particle during that transformation. \newline

We now apply this scheme to our case. For each probability density function $p(\mathbf{x})$ we may define, in analogy with classical mechanics, the vector quantity $\boldsymbol{\nu}(\mathbf{x}) = (\nu_1(\mathbf{x}),\nu_2(\mathbf{x}),\nu_3(\mathbf{x}))$ with components given by
\begin{equation}\label{nu}
	\nu_{i} := \lim_{h\to 0}\frac{d[p(\mathbf{x}+h\,\mathbf{e}_{i}),p(\mathbf{x})]}{h},
\end{equation}
where $d[p_1,p_2]$ is the distance between two probability density functions $p_1(\mathbf{x})$ and $p_2(\mathbf{x})$. Using the Jensen-Shannon distance \cite{2003-Endres,2003-Osterreicher}
\begin{equation}
	d_{JS}[p_{1},p_{2}] := \sqrt{ H\bigg[\frac{p_{1}+p_{2}}{2}\bigg]-\frac{1}{2}H[p_{1}]-\frac{1}{2}H[p_{2}]},
\end{equation}
where $H[p]$ is the differential entropy of the distribution $p(\mathbf{x})$, i.e.
\begin{equation}
\label{Shannon}
	H[p] := -\int p(\mathbf{x})\log p(\mathbf{x})d\mathbf{x}\,, 
\end{equation}
we have
\begin{equation}\label{nu2}
	\nu_{i}=\sqrt{\frac{1}{8}\int \frac{(\partial_{x_i}\,p(\mathbf{x}))^{2}}{p(\mathbf{x)}}d\mathbf{x}}.
\end{equation}
Now, if in analogy with the classical case we suppose that the energy needed to change the distribution function of a particle from $p_{1}=p$ to $p_{2}=p+dp$ is proportional to $\boldsymbol{\nu}\cdot d\boldsymbol{\nu}$, then we can assign a new kind of energy to the system that is proportional $\boldsymbol{\nu}^{2} = \boldsymbol{\nu} \cdot \boldsymbol{\nu}$ (like the kinetic energy, which is proportional to $\mathbf{v}^{2}$). In particular,
\begin{equation}\label{inf-eng}
	E_{\mathrm{inf}}=\gamma \boldsymbol{\nu}^{2}=\frac{\gamma}{8}\int \bigg(\frac{\nabla p(\mathbf{x})}{p(\mathbf{x})}\bigg)^{2}p(\mathbf{x})d\mathbf{x} ,
\end{equation}
where $\gamma$ is some constant with the dimension of an energy multiplied by a volume square. This quantity can be understood as the average energy needed to change the informational degree of freedom of a particle. Note that by setting $\gamma=\hbar^2/m$ we see that (\ref{inf-eng}) is equal to $U_q$. 
\newline


\section{Thermodynamics of the Quantum Potential}\label{sec3}

From the argument presented in the previous section, we saw that there are good reasons to suppose that the term $U_q$ is the average energy contribution related to the informational degree of freedom of a quantum system. In this section we develop a simple equilibrium thermodynamics for this degree of freedom. Consider the equation \eqref{QP}. From the discussion done in the previous section, we may consider the quantity
\begin{equation}\label{energyinf}
E[p,\nabla p]:=\frac{\hbar^{2}}{8m}\bigg(\frac{\nabla p}{p}\bigg)^{2},
\end{equation}
as the energy associated to the informational degree of freedom in the point $\mathbf{x}$ of configuration space. Let us now compute how this energy varies when the probability distribution $p(\mathbf{x})$ is varied to $p(\mathbf{x})+\delta p(\mathbf{x})$. We have
\begin{equation*}
\begin{split}
	\delta E[p,\nabla p ] &= E[p+\delta p, \nabla p +\nabla \delta p] - E[p,\nabla p] \\
								   &= 2 E[p,\nabla p ] \left\{ \frac{\sum_{i = 1}^3 \partial_{x_i}p\partial_{x_i} \delta p }{\sum_{i=1}^3 (\partial_{x_i}p)^2 }- \frac{\delta p}{p} \right\}.
\end{split}
\end{equation*}
 Calling $\| \nabla p \|$ the usual Euclidean norm in $\Rea^3$ of the vector $\nabla p$, one can easily obtain that
\begin{equation*}
	\delta \|\nabla p \| = \frac{1}{2 \sqrt{(\nabla p)^2}}\delta (\nabla p)^2 =  \frac{\sum_{i = 1}^3 \partial_{x_i}p\partial_{x_i} \delta p }{\sqrt{\sum_{i=1}^3 (\partial_{x_i}p)^2} }.
\end{equation*}
Hence, the variation of the energy $\delta E[p,\nabla p]$ is given by
\begin{equation}\label{pre-firstlaw}
		\delta E[p,\nabla p ] = 2 E[p,\nabla p] \left\{ \frac{\delta \|\nabla p \| }{\| \nabla p \|} - \frac{\delta p}{p} \right\}.
\end{equation}
At this point, in analogy with classical thermodynamics, one may impose that at \emph{equilibrium} an equipartition theorem holds. More precisely, at equilibrium $E[p,\nabla p] $ can be interpreted as the average energy of the informational degree of freedom in the point $\mathbf{x}$, and so
\begin{equation}\label{equipartition-theorem}
	E[p,\nabla p] = \frac{1}{2} k_b \tilde{T}
\end{equation}
where $k_b$ is the Boltzmann constant and $\tilde{T}$ plays the role of a temperature.  A more detailed discussion on the temperature introduced here, will be done in section \ref{discussion-section}. For the moment we observe that $\tilde{T}$ would depend on $\mathbf{x}$. If we use \eqref{equipartition-theorem} in \eqref{pre-firstlaw}, then the variation of the energy $\delta E[p,\nabla p]$ takes the suggestive form
\begin{equation}\label{firstlaw}
	\delta E[p,\nabla p ] = - k_b\tilde{T}\frac{\delta p}{p} + k_b\tilde{T}\frac{\delta \|\nabla p \| }{\| \nabla p \|} .
\end{equation}
This last relation resembles the \emph{first law of thermodynamics}. Such identification is possible if we choose to treat $p$ and $\| \nabla p \|$ as two independent variables and interpreting the two terms of \eqref{firstlaw} as the heat and work. This will be discussed in what follows. 

We conclude by observing that the generalization of the argument presented in section \ref{sec2} and equation \eqref{firstlaw} to the $n$-particle case is also possible. However, care in this last case is required for the identification of heat and work. We shall do this in section \ref{sec:n-particle}.

\subsection{Heat}\label{subsec:heat}

Let us first consider the first term in \eqref{firstlaw}. Assume, as is done in ordinary statistical mechanics, that the thermodynamical entropy associated to the particle with probability distribution $p(\mathbf{x})$ is given by
\begin{equation}
\mathcal{S}[p]= - k_b\int p(\mathbf{x})\log p(\mathbf{x})d\mathbf{x} = \int p(\mathbf{x}) S[p(\mathbf{x})]d\mathbf{x}
\end{equation} 
where we set
\begin{equation}\label{entropy}
S[p] :=-k_b\log p(\mathbf{x}).
\end{equation}
The quantity $S[p]$ can be treated as the \emph{entropy} associated to the particle at each point $\mathbf{x}$ of configuration space. Note that if $k_b = 1/2$, $S[p](\mathbf{x})$ coincides with the quantum entropy used in \cite{2009-Sbitnev}, to derive the quantum potential Q directly.
 
With this definition of entropy, the infinitesimal heat exchange when the temperature associated to the particle is $\tilde{T}$ would be  
\begin{equation}\label{diffheat}
\delta Q= \tilde{T}\delta S[p]=-k_b \tilde{T} \frac{\delta p}{p}.
\end{equation}
This justifies the interpretation of the first term of \eqref{firstlaw} as the heat contribution to the particle energy due to a variation of the probability distribution of that particle.

\subsection{Work}\label{subsec:work}

Since $\delta p$ is a result of the heat exchange, one can consider $\delta \|\nabla p\|$ as a result of work exchange
\begin{equation}\label{work}
\delta W=A\,\delta \|\nabla p\|,
\end{equation}
where $A$ is  the average value of some thermodynamical quantities denoted by $\mathcal{A}$.

In statistical mechanics, for a physical system at equilibrium with a thermal bath at a given temperature $T$, the probability that the quantity $\mathcal{A}$ has the value $a$ is given by the Boltzmann distribution, which is
\begin{equation}
P(\mathcal{A}=a)=\frac{1}{Z}\exp\left(-\frac{H[a]}{k_bT}\right),
\end{equation}
where $H$ is the part of the microscopic Hamiltonian of the system that is related to the quantity $\mathcal{A}$, and $Z$ is the partition function
\begin{equation*}
Z:=\int_{0}^{\infty}\exp\left(-\frac{H[a]}{k_bT}\right)da.
\end{equation*}
Using Eq.~(\ref{work}) we could write
\begin{equation}
H[a]= a \|\nabla p\|,
\end{equation}
thus the average value of $\mathcal{A}$ when the equilbrium temperature is $\tilde{T}$ would be
\begin{equation*}
A=\int_{0}^{\infty} a P(\mathcal{A} = a)\,da=\frac{k_b\tilde{T}}{\|\nabla p\|}.
\end{equation*}
Hence the work exchange at equilibrium is given by
\begin{equation}\label{workfin}
	\delta W = \frac{k_b\tilde{T}}{\|\nabla p\|}\delta \|\nabla p \|.
\end{equation}
This discussion shows that, assuming an informational degree of freedom for a particle in thermodynamical equilibrium with associated temperature $\tilde{T}$, the second term of \eqref{firstlaw} can be interpreted as the infinitesimal work done on the particle when the probability distribution $p$ changes. 
\newline

\begin{remark}
From the above argument we see that the change in the value of the probability distribution $p$ can be treated as \emph{heat exchange} of the quantum system, and the change in the value of the modulus of its spatial derivative, $\|\nabla p\|$, can be treated as \emph{work exchange} of the quantum system. 
\end{remark} 

From the discussion done till now, we can summarize the proposed thermodynamical derivation of the quantum potential for a single particle as follow. The quantum potential can be considered as the average energy of the informational degree of freedom. In particular, one can defining an energy for the informational degree of freedom of the particle in each point $\mathbb{x}$. Under an hypothesis of thermodynamical equilibrium and by associating a temperature to this energy by the equipartition theorem, we can apply thermodynamical methods to derive its form. In fact by using (\ref{diffheat}) and (\ref{workfin}) as expressions for the heat and the work exchange respectively, one can derive $U_q$ just by integrating (\ref{firstlaw}) and then taking the average, up to some multiplicative constant. At this point using \eqref{QPformula}, the quantum potential $Q$ can be derived.


\section{$N$-particle case}\label{sec:n-particle}

We saw in the previous section how to define the heat and work associated to a single quantum particle. In this section we want to discuss the $N$-particle case. The first thing we have to observe is that the temperature defined by the relation \eqref{equipartition-theorem} does not depend on a single point of the physical space in which the particle exists. Indeed, inverting \eqref{equipartition-theorem} we have
\begin{equation}\label{temperatureN}
	\tilde{T} = \frac{\hbar^2}{4k_b m}\bigg(\frac{\nabla |\psi(\mathbf{x}_1,\cdots, \mathbf{x}_N,t)|^2}{|\psi(\mathbf{x}_1,\cdots, \mathbf{x}_N,t)|^2}\bigg)^{2},
\end{equation}
where $\nabla$ is now understood in $\Rea^{3N}$, showing that $\tilde{T} = \tilde{T}(\mathbf{x}_1,\cdots, \mathbf{x}_N)$. This means that the temperature associated to these informational degrees of freedom is a temperature in the \emph{configuration space} and not in real space. Hence $\tilde{T}$ cannot be considered as the temperature of something existing in the 3-dimensional space in which the particles move. We will come back on this point in section \ref{sec6}. 

Let us restrict our attention to the $2$-particle case only. The observations we can do in this specific case, generalize straightforwardly to an arbitrary number of particles. 
Let $p_{12}(\mathbf{x}_1,\mathbf{x}_2)$ be the probability distribution of the two particles. Suppose they are prepared in an independent way and that they do not interact. In this case we have
\begin{equation}\label{two-prob}
	p_{12}(\mathbf{x}_1,\mathbf{x}_2) = p_1(\mathbf{x}_1)p_2(\mathbf{x}_2).
\end{equation}
Using (\ref{two-prob}) we obtain
\begin{equation}
	\frac{\delta p_{12}}{p_{12}}=\frac{\delta p_{1}}{p_{1}}+\frac{\delta p_{2}}{p_{2}},
\end{equation}
which according to (\ref{diffheat}) implies that the heat exchange for two independent non-interacting particles is an additive quantity. But since
\begin{equation}
	\|\nabla p_{1,2}\|\neq \|\nabla p_{1}\| \cdot \|\nabla p_{2}\|	
\end{equation}
we obtain
\begin{equation}
	\frac{\delta \|\nabla p_{1,2}\|}{\|\nabla p_{1,2}\|}\neq \frac{\delta \|\nabla p_{1}\|}{\|\nabla p_{1}\|}+\frac{\delta \|\nabla p_{2}\|}{\|\nabla p_{2}\|}.
\end{equation}
This, according to \eqref{workfin}, implies that the (informational) work for two particles is not an additive quantity even if they are independent and non-interacting, in contrast with ordinary thermodynamics. To overcome this difficulty one may simply consider a two particle system as a single system in the configuration space $\Rea^6$. Doing that, the additivity of the heat and work is not anymore a problem, since we always deal with a single system. In this way the quantum potential $Q$ for two particles can be derived following the same method explained at the end of the previous section. Note that, this is in agreement with the fact the in 2-particle case, the temperature \eqref{temperatureN} characterizing the informational equilibrium must be thought to be in configuration space.

\section{The temperature $\tilde{T}$}\label{sec6}\label{discussion-section} 

The extension to the $N$-particle case is rather trivial. Indeed, it is enough to repeat the whole analysis done in the previous section on $\Rea^{3N}$, the configuration space of an $N$-particle system. In light of this extension, we are now in the position to discuss the meaning of the temperature $\tilde{T}$ and the thermodynamical/informational equilibrium used. Given an $N$-dimensional quantum system with wave function $\psi(\mathbf{x},t)$, the temperature $\tilde{T}_{\psi}$ can be defined via the relation \eqref{equipartition-theorem} as
\begin{equation}\label{temperature}
	\tilde{T}_{\psi} = \frac{\hbar^2}{4k_b m}\bigg(\frac{\nabla |\psi(\mathbf{x},t)|^2}{|\psi(\mathbf{x},t)|^2}\bigg)^{2} = \frac{\hbar^2}{k_b m}\bigg(\frac{\nabla |\psi(\mathbf{x},t)|}{|\psi(\mathbf{x},t)|}\bigg)^{2}.
\end{equation} 
As already noted, this temperature does not depend on the intensity of $\psi(\mathbf{x},t)$, but on how it varies in space, a feature shared with the quantum potential. We also observe that as $\hbar \rightarrow 0$, this temperature vanish. This is in agreement with known results: in classical mechanics, such additional informational degree of freedom is  not taken into account. In what follow we study the behavior of the temperature (\ref{temperature}) by giving two examples.

\subsection*{Example 1: Free Particle}

\begin{center}
\begin{figure}[!htb]
	\minipage{0.32\textwidth}
	\includegraphics[width=\linewidth]{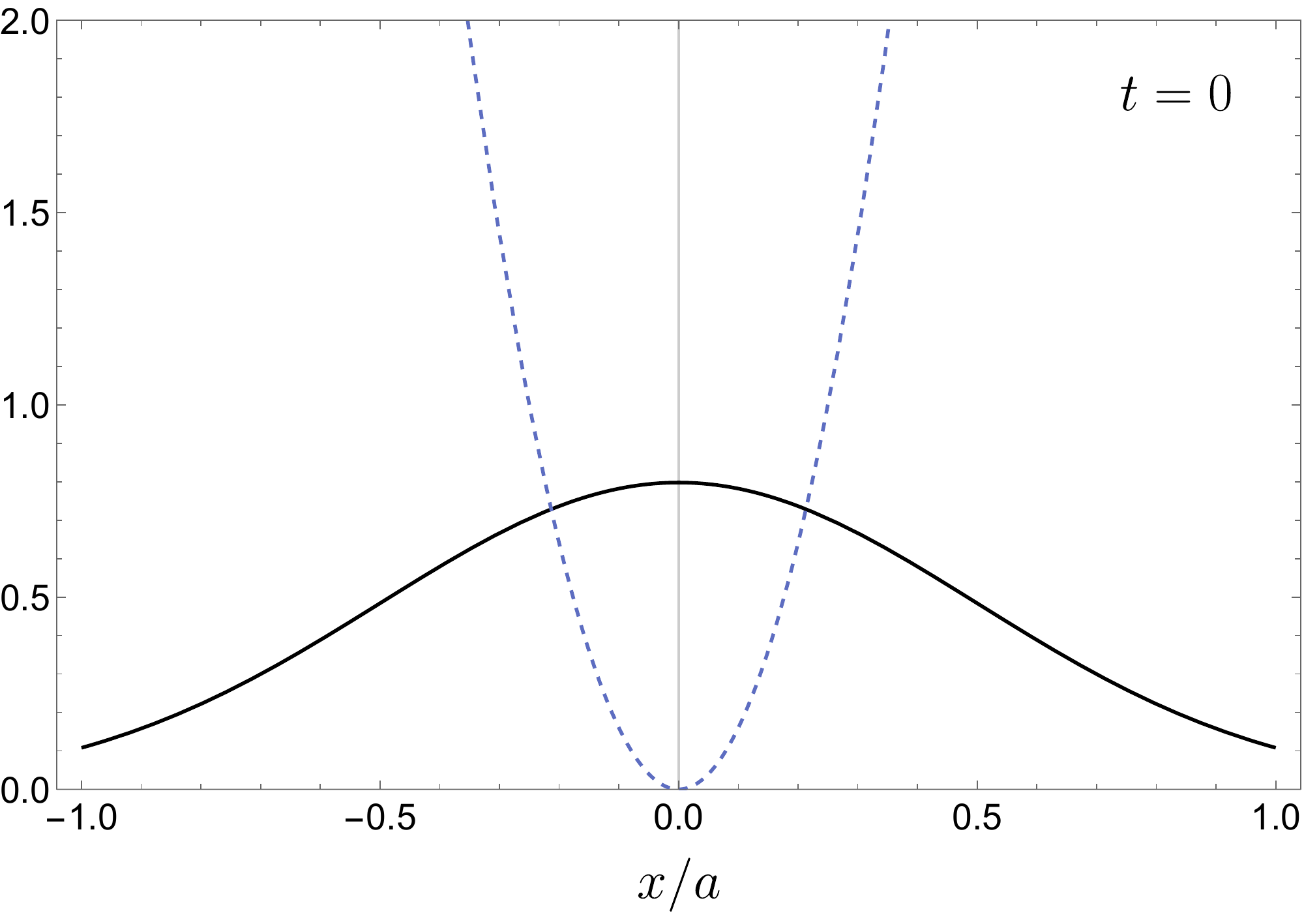}
	\endminipage\hfill
	\minipage{0.32\textwidth}
	\includegraphics[width=\linewidth]{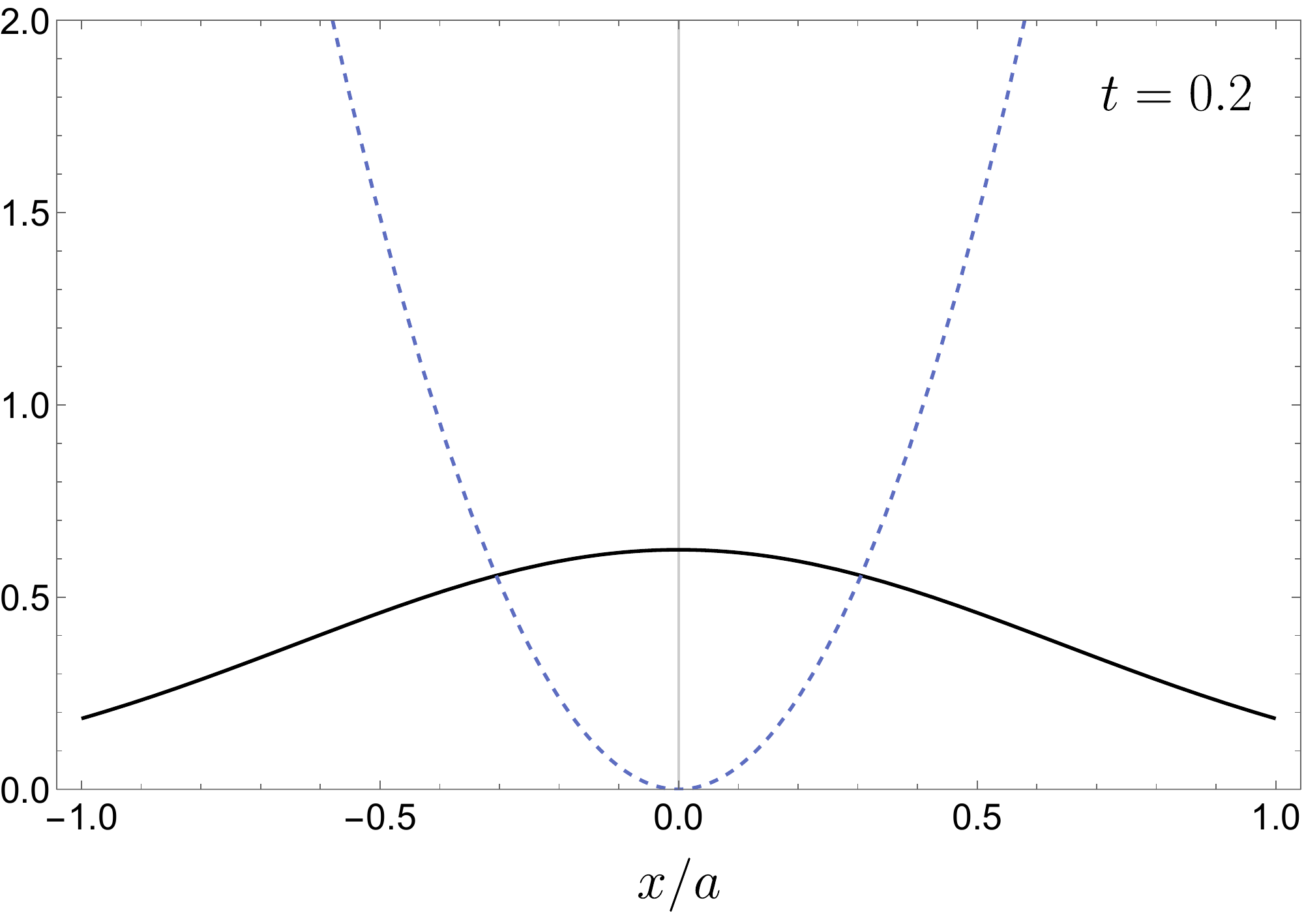}
	\endminipage\hfill
	\minipage{0.32\textwidth}%
	\includegraphics[width=\linewidth]{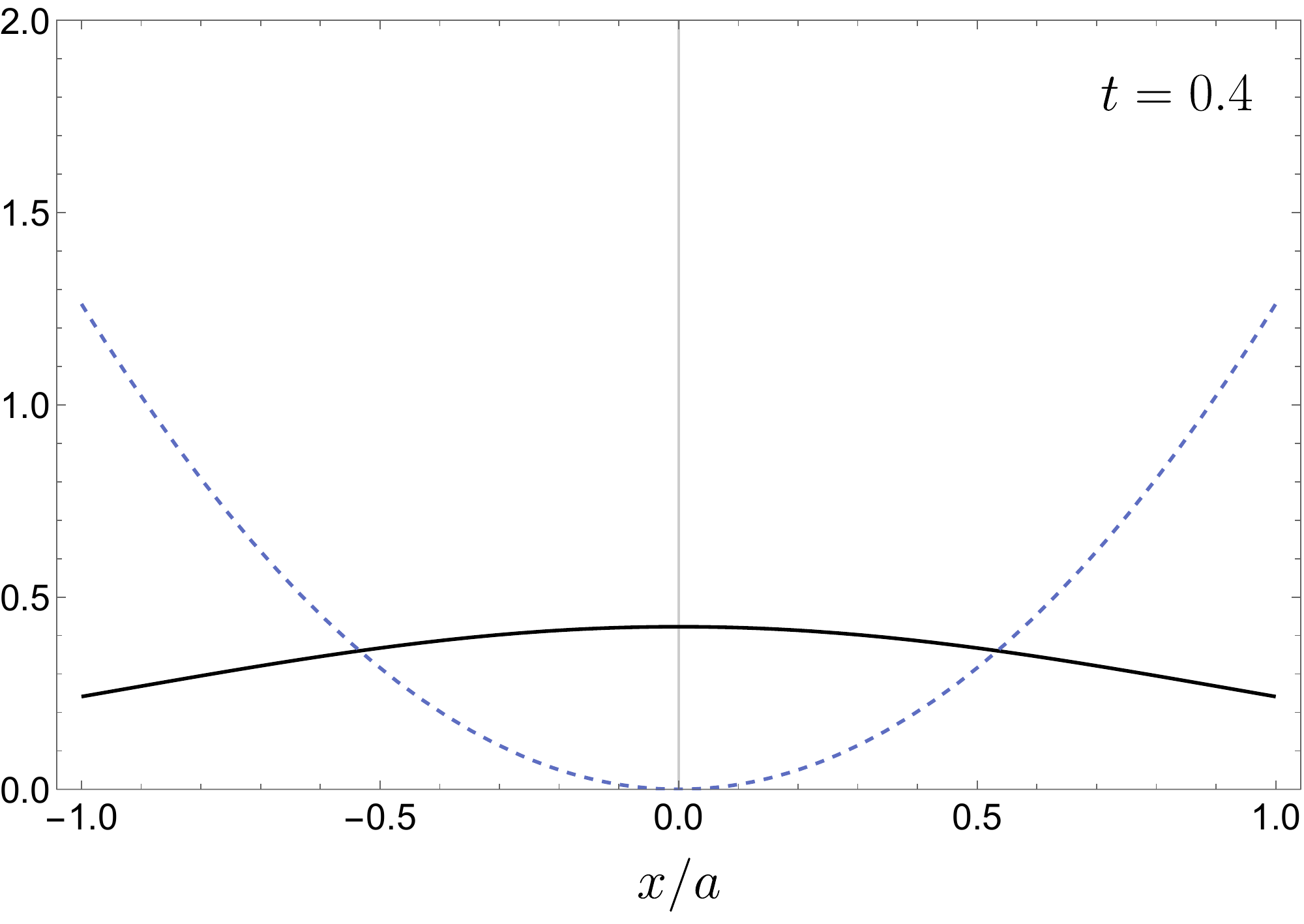}
	\endminipage
	\caption{Time evolution of the temperature field of a free particle with initial Gaussian probability distribution. The dashed and solid curves show respectively the temperature ($\times\frac{4k_{b}ma^{2}}{\hbar^{2}}$) and probability fields.}
	\label{FIG-1}
\end{figure}
\end{center}

For a free particle (i.e. $V=0$) \textcolor{blue}{ in 1-D} with initial wave function
\begin{equation}
	\psi(x,0)=\bigg(\frac{2}{\pi a^2}\bigg)^{1/4}\exp[-x^{2}/a^{2}],
\end{equation}
by solving the Schr\"{o}dinger equation (\ref{Shro-Eq}) we obtain
\begin{equation}
	\psi(x,t)=\bigg(\frac{2}{\pi a^2 (1+16t^{2})}\bigg)^{1/4}\exp[-x^{2}/a^{2}(1+16t^{2})],
\end{equation}
thus the temperature is
\begin{equation}
	\tilde{T}(x,t)=\frac{4\hbar^2}{k_{b}ma^{4}}\frac{x^{2}}{1+16t^{2}}.
\end{equation} 
Figure (\ref{FIG-1}) shows the time evolution of the temperature field. As we see, eventually all the points tend to thermalize to a common temperature. This is what one expects from an unconstrained thermodynamical system.

\subsection*{Example 2 : Particle in a Box}

\begin{center}
\begin{figure}[!htb]
	\minipage{0.32\textwidth}
	\includegraphics[width=\linewidth]{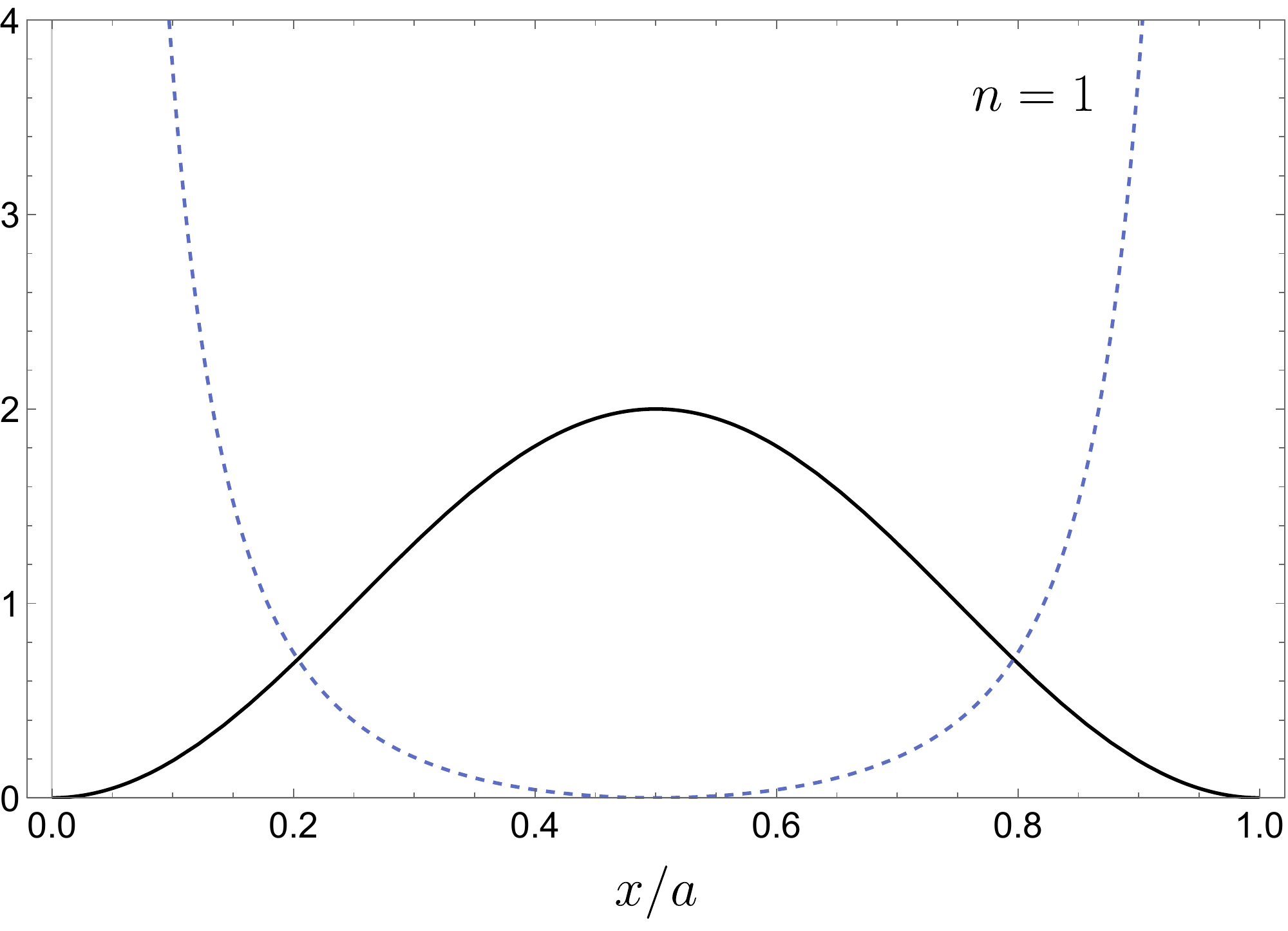}
	\endminipage
	\hspace{1cm}
	\minipage{0.32\textwidth}
	\includegraphics[width=\linewidth]{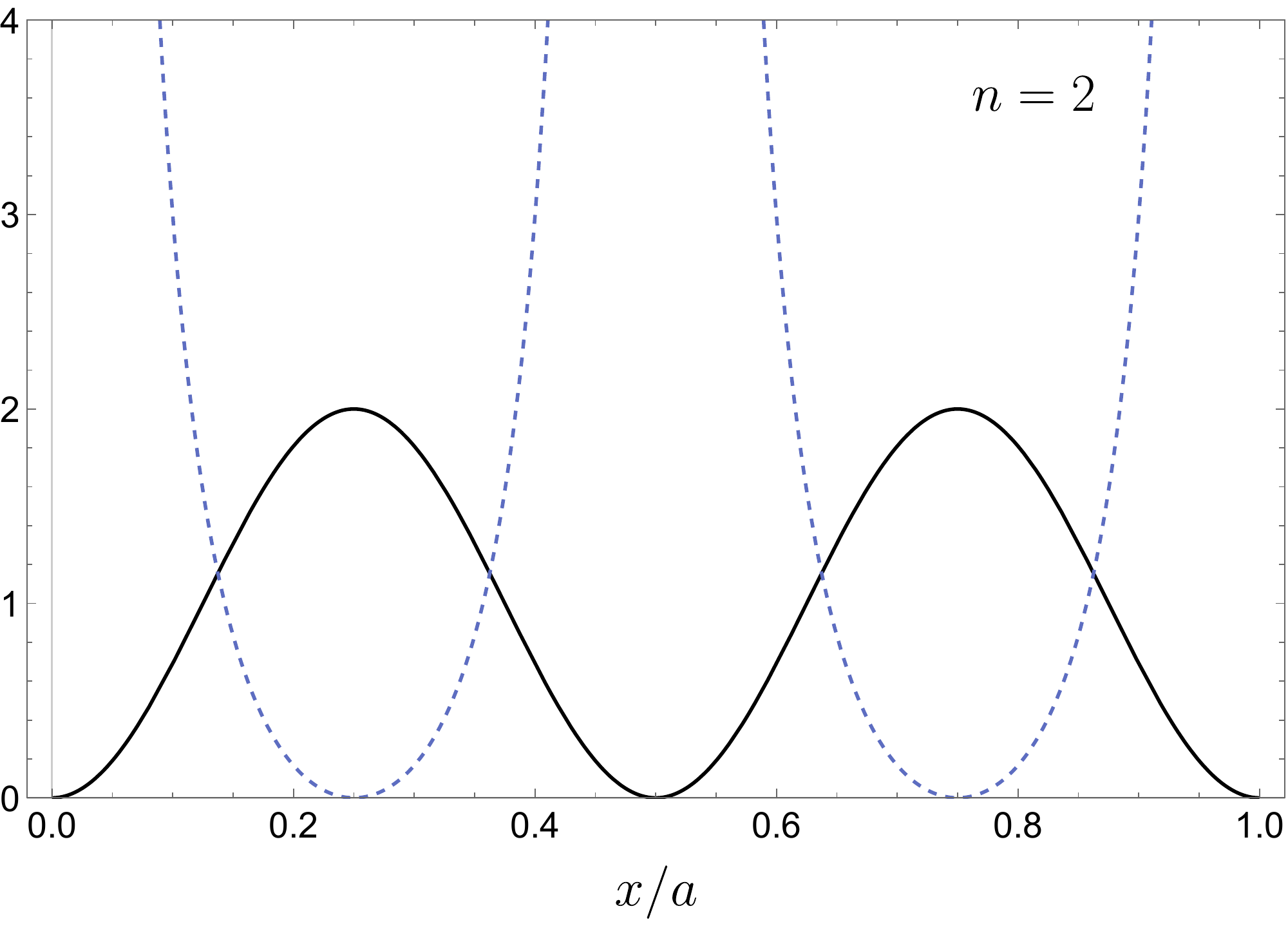}
	\endminipage\hfill
	\caption{Temperature field of the ground state and of the first excited state of a particle in a box. The dashed and solid curves show respectively the temperature ($\times\frac{4k_{b}ma^{2}10^{2}}{\hbar^{2}}$) and probability fields.}
	\label{FIG-2}
\end{figure}
\end{center}

For constrained thermodynamical systems one does not expect that in the steady state all points have the same temperature. In fact in such cases there will be always a temperature gradient in the final state of the system. A particle in a box is the simplest example of a constrained system. For a particle in a \textcolor{blue}{$1$-D} box of length $a$, the steady states of the system are
\begin{equation}
	\psi_{n}(x)=\sqrt{\frac{2}{a}} \sin\bigg(\frac{n\pi x}{a}\bigg),
\end{equation}
thus the steady state temperature fields would be
\begin{equation}
	\tilde{T}_{n}(x)=\frac{\hbar^{2}n^{2}\pi^{2}}{k_{b}ma^{2}}\cot^{2}\bigg(\frac{n\pi x}{a}\bigg).
\end{equation}
Figure (\ref{FIG-2}) shows the temperature field of the ground state and the first exited state of a particle in a box.

As we see in these two examples, where the probability is maximum/zero the temperature is zero/infinite. Thus, in these two cases, the low/high temperature correspond to high/low probability regions. More generally, from \eqref{temperature} we can see that the temperature vanishes in all the regions of configuration space in which the modulus of the wave function is constant, i.e. $\nabla|\psi(\mathbf{x})| = 0$, which corresponds to all the points in which $Q$ vanish (see \eqref{quantumpotential}).


\section{Conclusion.}

In this work we showed how, by means of simple equilibrium thermodynamical considerations it is possible to derive the quantum potential. In this way, we are able to justify the form of the Hamiltonian functional \eqref{Hamiltonian}, from which one can derive the equations of motion \eqref{continuity} and \eqref{Hamil-Jacob}. This derivation is based first, on the existence of an additional informational degree of freedom for any quantum system, second, on the association of a temperature to this quantum system, and third, 
on the validity of an equipartition theorem for this new degree of freedom. By studying some examples, we have seen that the introduced temperature behaves as one expected from thermodynamics.

We observed that temperature is defined in the configuration space of the system and  this precludes the interpretation of this quantity as a temperature field in the ordinary 3D space.  However an interesting meaning can be attached to this quantity. In Bohmian mechanics, the basic assumption that the initial probability distribution of the positions of particles is given by the square modulus of the associated wave function, i.e.
\begin{equation*}
	\rho(\mathbf{x},0) = |\psi(\mathbf{x},0)|^2,
\end{equation*}
is called \emph{quantum equilibrium hypothesis}. When this condition does not hold, we speak of  \emph{quantum non-equilibrium}. It was showed in \cite{1991-Valentini1,1991-Valentini2} that by some assumptions, when starting out of quantum equilibrium one rapidly converges toward quantum equilibrium; however this is not universally accepted \cite{1992-Durr2}. Since our temperature is defined once the probability distribution of the system $\rho(\mathbf{x},t)$ is exactly given by the square modulus of a wave function $\psi(\mathbf{x},t)$, this temperature can be thought as the \emph{temperature of the quantum equilibrium}. 

To conclude we observe that starting from equilibrium thermodynamics we have recovered quantum equilibrium. It is tempting to study what happens if one starts from non-equilibrium thermodynamics. Is it possible to derive some non-equilibrium quantum theory by using this method? Moreover, extending the method proposed here to the non-equilibrium case, may also be relevant for the justification of the Wallstrom condition \cite{1994-Wallstrom}, as suggested in \cite{2011-Schmelzer} where fluctuations from quantum equilibrium seem to be responsible for such condition.


\begin{thebibliography}{100}

\bibitem{1902-Gibbs} Gibbs, J. W.: "The Principles of Statistical Mechanics", Elementary Principles in Statistical Mechanics,   New York: Charles Scribner's Sons (1902).

\bibitem{1927-Madelung} Madelung, E.: "Quantentheorie in hydrodynamischer form", Zeit. f. Phys., 40, 322 (1927).

\bibitem{1938-Tolman} Tolman, R. C.: "The Principles of Statistical Mechanics", Dover Publications (1938).

\bibitem{1945-Rao} Rao, C. R.: "Information and the accuracy attainable in the estimation of statistical parameters", Bulletin of the Calcutta Mathematical Society 37, 81-89 (1945).%

\bibitem{1946-Cramer} Cram\'{e}r, H.: "Mathematical Methods of Statistics", Mathematical Methods of Statistics, Princeton, NJ: Princeton Univ. Press (1946).

\bibitem{1948-Shannon} Shannon, C. E.: "A Mathematical Theory of Communication", Bell Syst. Tech. J. 27, 379-423 (1948).

\bibitem{1952-Bohm1}  Bohm, D.: "A new suggested interpretation of quantum theory in terms of hidden variables. Part I.", Phys. Rev. 85, 166–179 (1952).

\bibitem{1952-Bohm2}  Bohm, D.: "A new suggested interpretation of quantum theory in terms of hidden variables. Part II.", Phys. Rev. 85, 180–193 (1952).

\bibitem{1966-Nelson} Nelson, E.: "Derivation of the Schrödinger Equation from Newtonian Mechanics.", Phys. Rev. 150, 1079 (1966).

\bibitem{1979-Philippidis} Philippidis, C., Dewdney, C., Hiley, B. J.: "Quantum interference and the quantum potential." Il Nuovo Cimento B, 52(1), 15-28 (1979). 

\bibitem{1982-Dewdney} Dewdney, C., Hiley, B. J.: "A quantum potential description of one-dimensional time-dependent scattering from square barriers and square wells." Found. Phys., 12(1), 27-48 (1982).

\bibitem{1983-Guerra} Guerra, F., Morato, L. M.: "Quantization of dynamical systems and stochastic control theory." Phys. Rev. D, 27(8), 1774 (1983).

\bibitem{1985-Callen} Callen, H.: "Thermodynamics and an introduction to thermostatistics", John Wiley \& Sons (1985)

\bibitem{1988-Guerra} Guerra, F.: "Stochastic variational principles in quantum mechanics.", Ann. Inst. Henry Poincare, 49, 315 (1988).

\bibitem{1989-Frieden} Frieden, B. R.: "Fisher information as the basis for the Schrödinger wave equation ",  Am. J. Phys. 57, 1004–1008 (1989).

\bibitem{1991-Valentini1} Valentini, A.: "Signal-locality, uncertainty, and the subquantum H-theorem. I.",  Phys.Lett. A, 156(1-2), 5-11, (1991).

\bibitem{1991-Valentini2} Valentini, A.: "Signal-locality, uncertainty, and the subquantum H-theorem. II.",  Physics Letters A, 158(1-2), 1-8, (1991).

\bibitem{1992-Durr} Dürr, D., Goldstein, S., Zanghi, N.: "Quantum mechanics, randomness, and deterministic reality.", Phys. Lett. A, 172(1-2), 6-12 (1992). 

\bibitem{1992-Durr2}Dürr, D., Goldstein, S., Zanghi, N.: "Quantum equilibrium and the origin of absolute uncertainty". J. Stat. Phys., 67(5-6), 843-907 (1992). 

\bibitem{1994-Wallstrom} Wallstrom, T. C.: "Inequivalence between the Schr\"{o}dinger equation and the Madelung hydrodynamic equations.", Phys. Rev. A 49, 1613-1617 (1994).

\bibitem{1998-Reginatto} Reginatto, M.: "Derivation of the equations of nonrelativistic quantum mechanics using the principle of minimum Fisher information ", Phys. Rev. A \textbf{58}, 1775 (1998).

\bibitem{1998-Recami} Recami, E., Salesi, G.: "Kinematics and hydrodynamics of spinning particles.", Phys. Rev. A, 57(1), 98, (1998).

\bibitem{1925-Fisher} Fisher, R.: "Theory of Statistical Estimation.", Math. Proc. Camb. Philos. Soc. , 22(5), 700-725, (1925).

\bibitem{2000-Luo} Luo, S.: "Quantum Fisher Information and Uncertainty Relations", Lett. Math. Phys.  53, 243 (2000).

\bibitem{2007-Cohen} Cohen, E. R.; et al.: "IUPAC Green Book (3rd ed.)", Cambridge: IUPAC and RSC Publishing, (2007).

\bibitem{2003-Endres} Endres, D. M.,  Schindelin, J. E.:"A new metric for probability distributions.",  IEEE Trans. Inf. Theory. 49, 1858-1860 (2003).

\bibitem{2003-Osterreicher} Osterreicher, F., Vajda, I.: "A new class of metric divergences on probability spaces and its statistical applications.", Ann. Inst. Stat. Math. 55, 639-653 (2003).

\bibitem{2005-Taylor} Taylor, J.R.: "Classical Mechanics", University Science Books (2005).

\bibitem{2005-Hiley} Hiley, B. J.: "Non-commutative quantum geometry: a reappraisal of the Bohm approach to quantum theory.", In "Quo Vadis Quantum Mechanics?" (pp. 299-324). Springer, Berlin, Heidelberg, (2005).

\bibitem{2008-Grossing} Gr\"{o}ssing, G.: "The vacuum fluctuation theorem: Exact Schrödinger equation via nonequilibrium thermodynamics." Phys. Lett. A, 372(25), 4556-4563, (2008). 

\bibitem{2009-Grossing} Gr\"{o}ssing, G.: "On the thermodynamic origin of the quantum potential.", Phys.A, 388(6), 811-823, (2009). 

\bibitem{2009-Durr} Dürr, D., Teufel, S.: "Bohmian mechanics: the physics and mathematics of quantum theory." Springer Science \& Business Media (2009).

\bibitem{2009-Sbitnev} Sbitnev V. I.: "Bohmian trajectories and the path integral paradigm: complexified Lagrangian mechanics.", Int. J. Bifurcation Chaos 19.07, 2335-2346 (2009).

\bibitem{2010-Carroll} Carroll, R. W.: "On the emergence theme of physics.", World Scientific, (2010).

\bibitem{2010-Hiley} Hiley, B. J., Callaghan, R. E.: "The Clifford algebra approach to quantum mechanics A: The Schrödinger and Pauli particles." arXiv preprint arXiv:1011.4031,  (2010).

\bibitem{2012-Fiscaletti} Fiscaletti, D.: "The geometrodynamic nature of the quantum potential.",Ukran. Phys. J., 57, N. 5, 561-573 (2012). 

\bibitem{2013-Atiq} Atiq, M., Karamian, M., Golshani, M.: "A Quasi-Newtonian approach to Bohmian mechanics i: Quantum potential.", Ann. Fond. Louis de Broglie, Vol. 34, N. 1, arXiv:1311.6497, (2013).

\bibitem{2014-Licata} Licata, I., Fiscaletti, D.: "Quantum potential: Physics, geometry and algebra.", New York: Springer, (2014). 

\bibitem{2011-Schmelzer} Schmelzer I.: "An answer to the Wallstrom objection against Nelsonian stochastics", arXiv:1101.5774v3 (2011).



\end{thebibliography}
\end{document}